\documentclass[useAMS]{mn2e}
\usepackage{amssymb}
\usepackage{graphicx}
\usepackage{psfrag}
\usepackage{times,mathptm,mathrsfs,txfonts}
\makeatletter

\begin{document}

\input epsf

\title{Encounters between spherical galaxies II: \\ systems with dark halo}
\author[A.C. Gonz\'alez-Garc\'{\i}a \& T.S. van Albada]{A.C. Gonz\'alez-Garc\'{\i}a$^1$$^,$$^2$ and T.S. van Albada$^1$ \\
$^1$ Kapteyn Astronomical Institute,P.O. BOX 800, 9700 AV Groningen, The Netherlands\\
$^2$ Instituto de Astrof\'{\i}sica de Canarias, V\'{\i}a L\'actea s/n, La Laguna, 38200, Spain}

\maketitle

\begin{abstract}
We performe N-body simulations of encounters between spherical systems surrounded by a spherical halo. Following a preceding paper with a similar aim, the initial systems include a spherical Jaffe model for the luminous matter and a Hernquist model for the halo. The merger remnants from this sample are mainly slowly rotating, prolate spheroids with a radially anisotropic velocity distribution. The results are compared with real-life ellipticals and with the models without halo in paper I. We argue that elliptical galaxies with evidence of dark matter could be formed in the field via a merger of spheroids surrounded by a dark matter halo, while ellipticals with no evidence of dark matter might be formed via a merger of two spheroids in a cluster.

\end{abstract}


\section{Introduction}

In the preceding paper (Gonz\'alez-Garc\'{\i}a \& van Albada, 2005, hereafter Paper I), encounters of one-component spherical isotropic systems were studied. Several initial conditions were investigated systematically in order to link the end products of the simulations with real-life elliptical galaxies. Both mergers and non-mergers were produced and analysed. From that study we conclude that various properties of elliptical galaxies can be explained by interactions between spheroids.
In particular, ellipticity, rotation, and boxiness-disciness can be attributed to the initial amount of angular momentum in the collision orbit. The mass ratio of the progenitor galaxies and the orbital energy affect mainly the final outcome of the interaction (i.e. whether it is a merger or not) and the morphology of the merger remnant.

In the present paper we study how these results are modified when a dark halo is present.

The amount of dark matter in elliptical galaxies is still controversial. 
The lack of HI in elliptical galaxies complicates this issue and a number of alternative ways have been proposed to study the amount of dark matter (cf. gravitational lensing: Keeton, Kochanek \& Falco 1998; X-ray haloes: Matsushita et al.  1998; planetary nebula: Romanowsky et al.  2001; stellar kinematics: Saglia, Bertin \& Stiavelli  1992, Kronawitter et al. 2000). These studies show that there is evidence for the presence of dark haloes, although dark matter does not seem to be important in the inner ($R< R_{\rm e}$) parts of elliptical galaxies.

Saglia, Bender \& Dressler (1993), and more recently Capaccioli, Napolitano \& Arnaboldi (2002), argue that possibly two populations of ellipticals exist. One would be dark matter dominated while in the other a diffuse halo, or perhaps no dark matter at all, would be present.

Traditionally merger simulations involving a dark matter component have dealt with mergers between disc galaxies (see Barnes 1998 for a review). Simulations involving spherical systems do not usually deal with two-component models (for a review see Paper I). In recent years some effort has been devoted to modeling such encounters with the purpose to study the Fundamental Plane of elliptical galaxies (Dantas et al.  2003, Gonz\'alez-Garc\'{\i}a 2003, chapter 5) but little attention has been given to the characteristics of the remnant of those encounters.

In the present paper we study a sample of merger simulations of spherical systems embedded in a halo. These simulations do not intend to cover the entire initial parameter space. Rather they aim at a comparison with the no-halo merger remnants of Paper I and with observations.

\section{Models}

\subsection{Initial conditions}

In constructing the initial systems we follow the same strategy as in Paper I, i.e., we use Jeans' theorem and Eddington's formula to relate the distribution function (DF) to the potential of our systems.

The systems include a luminous bulge, representing the distribution of light in elliptical galaxies, and a dark matter halo. 
We use the Jaffe (1983) sphere used in Paper I, to model the luminous matter, and a Hernquist (1990a) sphere for the dark halo.
Jaffe and Hernquist models are two examples of potential-density pairs with an analytical solution for the distribution function. In fact they are part of a larger family of models studied by Dehnen (1993).

The relevant quantities are the following:
The potential for a Jaffe model is:

\begin{equation}
	\phi_{\rm L}(r) = \frac{GM_{\rm L}}{r_{\rm J}} \ln \left(\frac{r}{r+r_{\rm J}}\right),
\end{equation}
where G is Newton's constant of gravity, $M_{\rm L}$ is the luminous mass of the system and $r_{\rm J}$ is the half mass radius of the luminous component. The corresponding mass density is:

\begin{equation}
	\rho_{\rm L}(r) = \left(\frac{M_{\rm L}}{4\pi r_{\rm J}^3}\right)\frac{r_{\rm J}^4}{r^2(r+r_{\rm J})^2}.
\end{equation}

The potential of the Hernquist model is:

\begin{equation}
	\phi_{\rm H}(r) = -\frac{GM_{\rm H}}{r+a},
\end{equation}
where $M_{\rm H}$ is the mass of the halo and $a$ is the scale length. The corresponding mass density is:

\begin{equation}
	\rho_{\rm H}(r) = \left(\frac{M_{\rm H}}{2\pi r}\right)\frac{a}{(r+a)^3}.
\end{equation}

The half-mass radius of the dark halo component is equal to $(1 + \sqrt{2}) a$.
Combining a Jaffe model with a Hernquist model we have a two-parameter family that depends on the ratio between the masses of the two systems $M_{\rm L}/M_{\rm H}$ and the ratio between the half-mass radii:

\begin{equation}
	r_{\rm L1/2}/r_{\rm H1/2} =\frac{r_{\rm J}}{(1+\sqrt{2})a}.
\end{equation}

To find the distribution function for the two components separately, Eddington's formula (Binney \& Tremaine, eq. 4-140b) has to be solved for the different densities in the potential generated by the total density, with

\begin{eqnarray}
\rho_{\rm T} & = & \rho_{\rm L}+\rho_{\rm H}, \\
\phi_{\rm T} & = & \phi_{\rm L}+\phi_{\rm H}.
\end{eqnarray}

Then the distribution function of the combined system is :

\begin{equation}
	f_{\rm T}(E) =f_{\rm L}(E)+f_{\rm H}(E).
\end{equation}

An algorithm yielding $f_{\rm L}(E)$ and $f_{\rm H}(E)$ along the two-parameter family was developed by Smulders \& Balcells (1995). The models used here were constructed using that algorithm.

We took our systems to be spherical and non-rotating, with an isotropic velocity distribution. In this regard they are similar to the set of spherical isotropic systems without halo studied in Paper I.

\begin{figure}
\begin{center}
\leavevmode
\hbox{%
\epsfxsize=7cm
\epsfbox{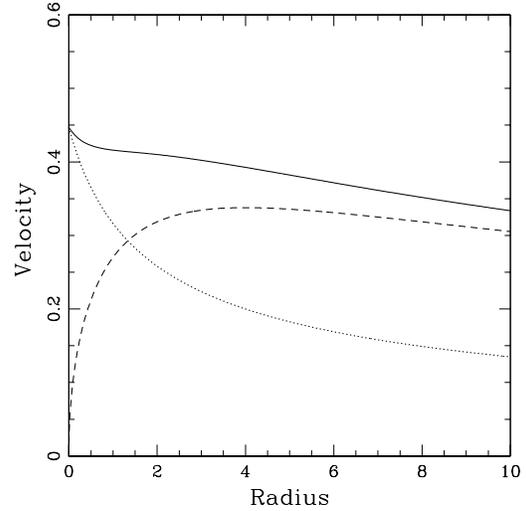}}
\caption{Circular velocity curves for the initial model. Parameters were chosen such that the circular velocity curve (top curve) is approximately flat. The dotted curve refers to the contribution by the Jaffe model and the dashed line to the Hernquist model. \label{fig:vel}}
\end{center}
\end{figure}

In our models  we have tried a range of $M_{\rm H}/M_{\rm L}$ values, and calculated the circular velocity curve for those realizations (bearing in mind that we use a Jaffe+Hernquist model) and aiming at a flat-topped curve for a significant radial range. After several trials we came to the conclusion that for the model with total mass $M_{\rm T}=1$ a value of $M_{\rm H}/M_{\rm L} = 9$ and a halo scale length $a=2$ (for $r_{\rm J}=1$) was the best choice for our purposes (see Figure \ref{fig:vel}).
A further scaling of this mass ratio was done following a scheme presented below (section 2.5). 

A cut-off radius has been imposed on the numerical implementation of this two-component model. The cut-off radius is equal to $10 \times r_{\rm J}$. The theoretical and experimental values of the half-mass radius are then no longer equal. A distinction between $r_J$ and $a$ (theoretical values) and $r_{\rm L1/2}$ and $r_{\rm H1/2}$ 
(computational values) will therefore be made. 

\subsection{Units}

The results in this paper are presented in non-dimensional units such that Newton's constant of gravity $G=1$, and the theoretical half-mass radius, $r_{\rm J}$, for the Jaffe model is also equal to one. Finally, the total mass of the initial system is also equal to unity. A possible set of units to scale our models to values for real galaxies, and allowing a comparison with observations, would be:

\begin{equation}
	        [M] = 10^{12} \; \rmn{M_{\odot}}, \\
\end{equation}

\begin{equation}
		[L] = r_{\rm J} = 5 \;\rmn{kpc},  \\
\end{equation}

\begin{equation}
		[T] =  5.27 \times 10^7 \;\rmn{yr}. \\
\end{equation}

With these, the unit for velocity is:

\begin{equation}
		[v] =  1000\; \rmn{km/s}. \\
\end{equation}

\subsection{Method}

As in Paper I, we used Hernquist's (1987, 1990b) version of the {\small TREECODE} on an Ultra-Sparc station. Softening was set to $1/8$ of the half-mass radius of the Jaffe core of the smallest galaxy ($r_{\rm L1/2} = 0.82$ and $\varepsilonup = 0.1$), the tolerance parameter was set to $\theta=0.8$, quadrupole terms were included in the force calculation and the time step was set to $1/100$ of the half-mass crossing time. We have used variable time steps, allowing refined calculations depending on the particle density. A typical run takes of the order of $5$ x $10^5$ seconds. 

\subsection{Stability of initial systems}

We have checked the stability of the various initial systems by letting them evolve for more than 10 crossing times of the smallest component (i.e. the Jaffe bulge in this case), and checking that the systems are stable at different radii. We do this for a system with 40000 particles in the Hernquist's halo and 10000 in the Jaffe component. 

\begin{figure}

\begin{center}
\leavevmode
\vbox{%
\epsfxsize=8cm
\epsfbox{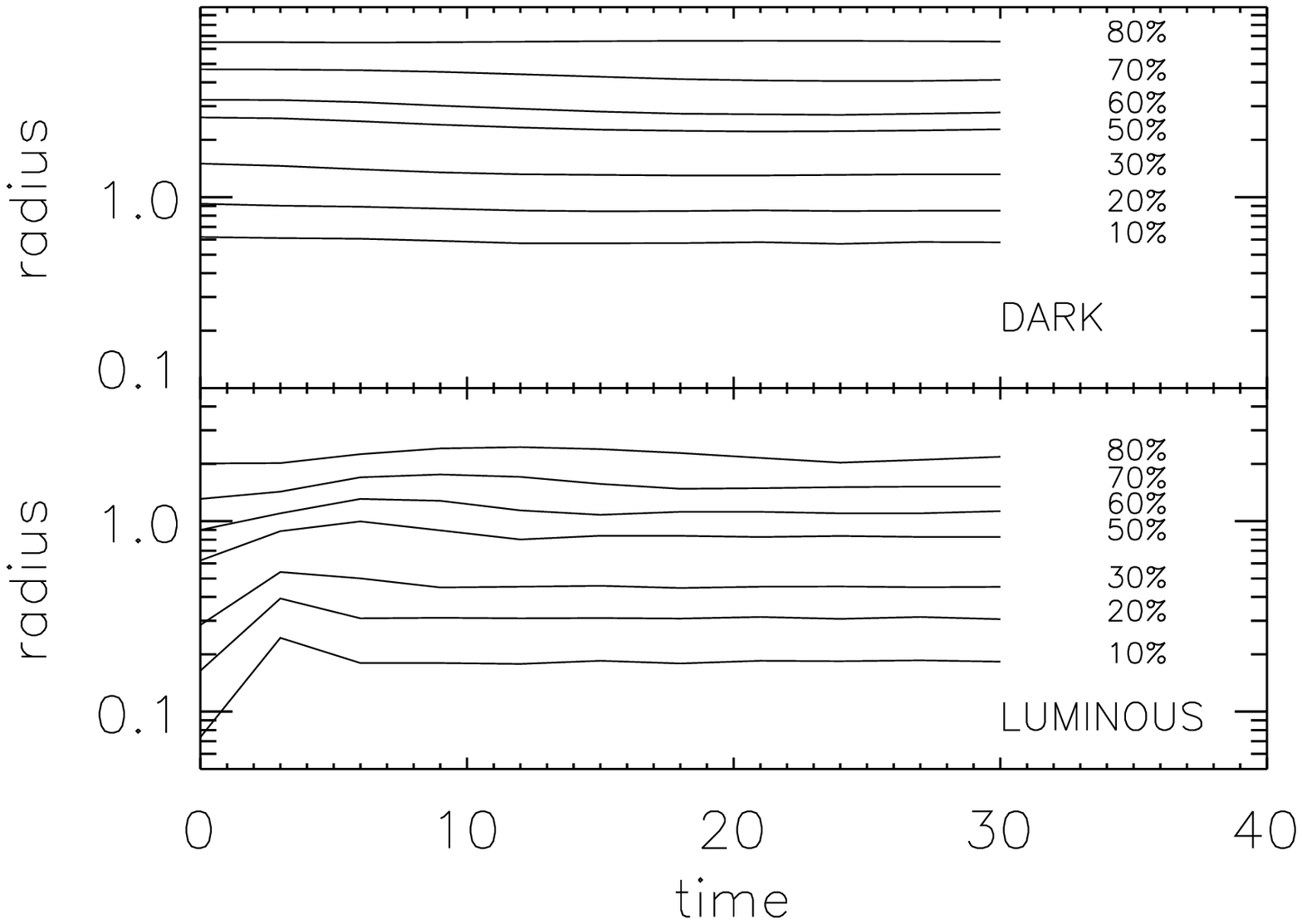}
\epsfxsize=8cm
\epsfbox{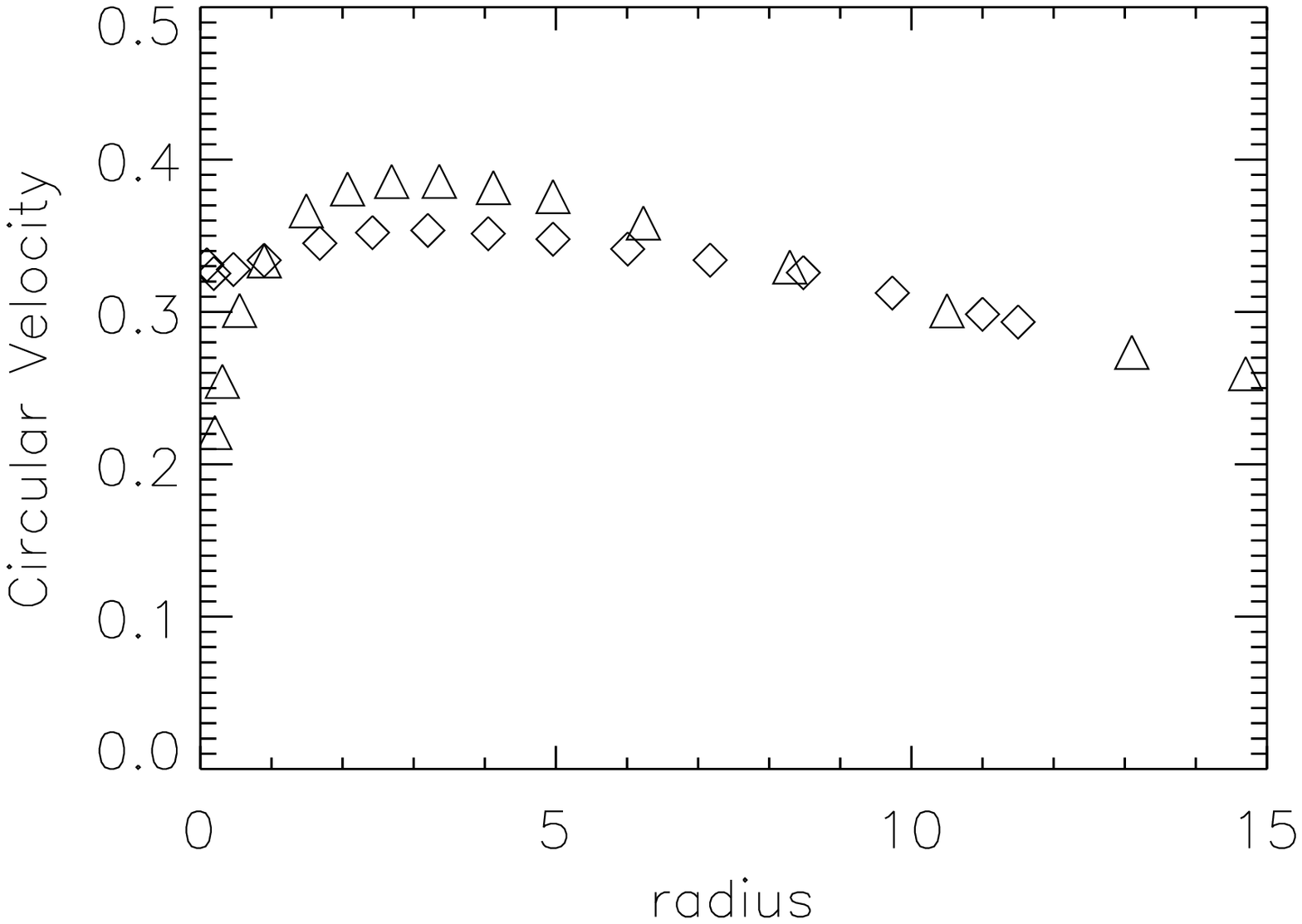}}
\caption{Top: Evolution of the radial mass distribution of the dark and luminous particles  for our bulge-halo system with $M_{\rm T}=0.5$. After a mild reorganization of the luminous component, the system is in equilibrium. Bottom: Circular velocity curve for the initial system (diamonds) and the relaxed one (triangles). \label{fig:vrH}}
\end{center}
\end{figure}

Figure \ref{fig:vrH} top shows that after some adjustment, the systems reach equilibrium. The expansion seen in the inner parts of the luminous component can probably be attributed to the softening introduced through the {\small TREECODE} in the force calculation. After this adjustment the circular velocity curve is still approximately flat (Figure \ref{fig:vrH}, bottom).

\subsection {Initial parameters}

We construct three different initial configurations, with masses $1/2$, $1$ and $2$. 

We have scaled the $M/L$ ratio inside these systems with the following assumptions:  the total mass of the system is given by the combined mass of the dark matter halo and luminous component. For the luminous part (Jaffe bulge) we take $M/L$ constant. 

Further, we assume that our systems would initially lie on the observed Fundamental Plane so the $M/L$ ratio of the combined system should follow an scaling law like $M/L \propto M^{\alpha}$ (Renzini \& Ciotti 1993). Following J\o rgensen, Franx \& Kj\ae rgaard (1996) we use:

\begin{equation}
 \log{L} \simeq 0.78 \log{M} + constant.
\label{eq:3.4.13}
\end{equation}

The scaling of the masses and radii between different mass models is a bit more tricky than that in Paper I, because now we want the luminous matter to be scaled following a relation between mass and luminosity similar to that found via the Fundamental Plane (eq. \ref{eq:3.4.13}) and the relation between mass and radius for different models given by Fish's (1964) law:

\begin{equation}
	\frac{R_1}{R_2}=\sqrt{\frac{M_1}{M_2}}
\label{eq:3.4.14}
\end{equation}

\begin{table}
\begin{center}

\caption{Initial parameters for the progenitor models.\label{tab2_0}}
\begin{tabular}{ccccc}
\hline
{\bf $M_{\rm T}$} &{\bf $M_{\rm L}$}& {\bf $r_{\rm J}$} & {\bf $M_{\rm H}$}& {\bf $a$} \\
\hline
0.5 & 0.0574 & 0.7576 & 0.4426 & 1.4025 \\
1 & 0.1 & 1 & 0.9 & 2 \\
2 & 0.1741 & 1.3195 & 1.8259 & 2.8487 \\
\hline  
\end{tabular}
\end{center}
\end{table}

\begin{table}
\begin{center}

\caption{Input parameters for the merger simulations. The columns give the model name, the mass ratio, impact parameter and orbital energy.\label{tab2_antes}}
\begin{tabular}{cccc}
\hline
{\bf Run} &{\bf $M_2:M_1$}& {\bf $Impact Par.$} & {\bf $E_{\rm orb}$} \\
\hline
$1h$ & 1:1 & 0 & 0 \\
$1o$ & 1:1 & 5 & 0 \\
$1g$ & 1:1 & 10 & 0 \\
$2h$ & 2:1 & 0 & 0 \\
$2o$ & 2:1 & 7.07 & 0 \\
$2g$ & 2:1 & 14.14 & 0 \\
$4h$ & 2:1/2 & 0 & 0 \\
$4o$ & 2:1/2 & 7.07 & 0 \\
$4g$ & 2:1/2 & 14.14 & 0 \\
\hline  
\end{tabular}
\end{center}
\end{table}


\begin{table}
\begin{center}
\caption{Properties of the merger remnants. Columns give the model number (1), the time where the run was stopped (2), the l.o.s. ellipticity at $R_{\rm e}$ from a point of view perpendicual to the inital angular momentum vector (3), the axis ratios $b/a$ (4), and $c/a$ (5), and the ratio between the maximum rotational velocity and the central velocity dispersion (6). \label{tab:halo}}
\begin{tabular}{cccccc}
\hline
{\bf Run} &{\bf $t_{\rm fin}$}& {\bf $\epsilon$}& { \bf $b/a$}& {\bf $ c/a$} & {\bf $V_{\rm max}/\sigma_{\rm o}$} \\
(1) & (2) & (3) & (4) & (5) & (6) \\
\hline
$1h$ & 144 & 0.621 & 0.676 &0.622  &0.157\\
$1o$ & 150 & 0.530 &0.684 &0.644  & 0.262\\
$1g$ & 225 & 0.564 &0.690 & 0.650 & 0.228\\
$2h$ & 200 & 0.593 &0.702 &0.669  & 0.175\\
$2o$ & 200 & 0.520 & 0.695 &0.681  & 0.175\\
$2g$ & 360 & 0.573 &0.772  &0.695  & 0.456\\
$4h$ & 225 & 0.359 & 0.757 &0.741  &0.162\\
$4o$ & 250& 0.265 &0.807  &0.756  & 0.173\\
$4g$ & 600 & 0.222 & 0.864 &0.788  &0.474\\
\hline
\end{tabular}
\end{center}
\end{table}

Using eq.~\ref{eq:3.4.13} we scaled the relation between luminous and total matter as follows:

\begin{equation}	
	\frac{M_{\rm T}^{4/5}}{M_{\rm L}} = constant,
\end{equation}
and we use Fish's law (eq. \ref{eq:3.4.14}) to scale the radii of the two components for the systems of different mass. The parameters of the initial systems are given in Table \ref{tab2_0}.

We restrict ourselves to parabolic orbits, so $E_{\rm orb}=0$ for all our simulations. The impact parameters are chosen such that the encounters  will lead to mergers.

To study the influence of orbital angular momentum on the merger remnant we have used three different impact parameters, head-on ($D=0$), and two off-axis: one with $D$ equal to half of the radius enclosing $99$ per cent  of the luminous mass of the bigger system, $D=r_{\rm L}/2$, and one with $D$ equal to that radius, $D=r_{\rm L}$, (this radius corresponds to $95$ per cent  of the total, luminous+dark, mass).

The resulting mass ratios, orbital energies and impact parameters are given in table \ref{tab2_antes}. The model numbers reflect the mass ratio and the type of orbit is indicated by a letter, with $h$ for head-on, $o$ for off set and $g$ for grazing encounters. This notation is similar to that used in Paper I, but now we do not include a letter for the orbital energy because it is always equal to zero.

In each simulation the two galaxies were initially placed at a distance between their edges equal to the radius of the smallest galaxy. In the simulations without halo described in Paper I twice this radius was used. For more than half of this distance tidal interaction was negligible however. To reduce computing time we therefore decided to use a separation between the centers equal to $2R_1+R_2$, where $R_1$ refers to the smallest system and $R_2$ to the largest. 

Each galaxy has $50000$ particles, $40000$ particles in the Hernquist realization of the dark halo, and $10000$ particles in the Jaffe luminous component.

\begin{table*}
\begin{center}
\begin{minipage}{120mm}
\caption{Morphological classification of the end products. The classification (column (2)) is indicative only; it refers to the value of the median when measuring the type from 100 random points of view. Columns (3) and (4) give a morphological description of each component.\label{jhmorf}}
\begin{tabular}{cccc}
\hline
{\bf Model}& {\bf Hubble Type}& {\bf Halo}&{\bf Luminous}\\
(1) & (2) & (3) & (4) \\
\hline
$1h$ &$E4-E5$ &$double\; lobe$ &$prolate,\; no \;rotation$\\
$1o$ &$E4$   &$symmetrical\;  tail$&$two\; tails,\; prolate-triaxial,\; no \; rot.$  \\
$1g$ &$E4-E5$  &$prominent\; double\; tail$&$triaxial,\; small \; rotation$  \\
$2h$ &$E5$  &$double\; asymmetrical \;lobe$&$prolate\;  + \; plume$ \\
$2o$ &$E3-4$  &$single \;tail$ &$nearly \; triaxial$\\
$2g$ &$E3-E4$ & $double\; asymmetric\; tail$&$triaxial,\; rotation$ \\
$4h$ &$E3-E4$  &$single\; lobe$&$prolate,\; shells$ \\
$4o$ & $E3$ &$several\; tails$ &$tails\; + \;shells$\\
$4g$ &$E3$  &$several\; tails$&$tails\; +\; shells \;(prom), \;rotation.$\\
\hline
\end{tabular}
\end{minipage}
\end{center}
\end{table*}

\begin{figure*}
\begin{center}
\leavevmode
\hbox{%
\epsfxsize=14cm
\epsfbox{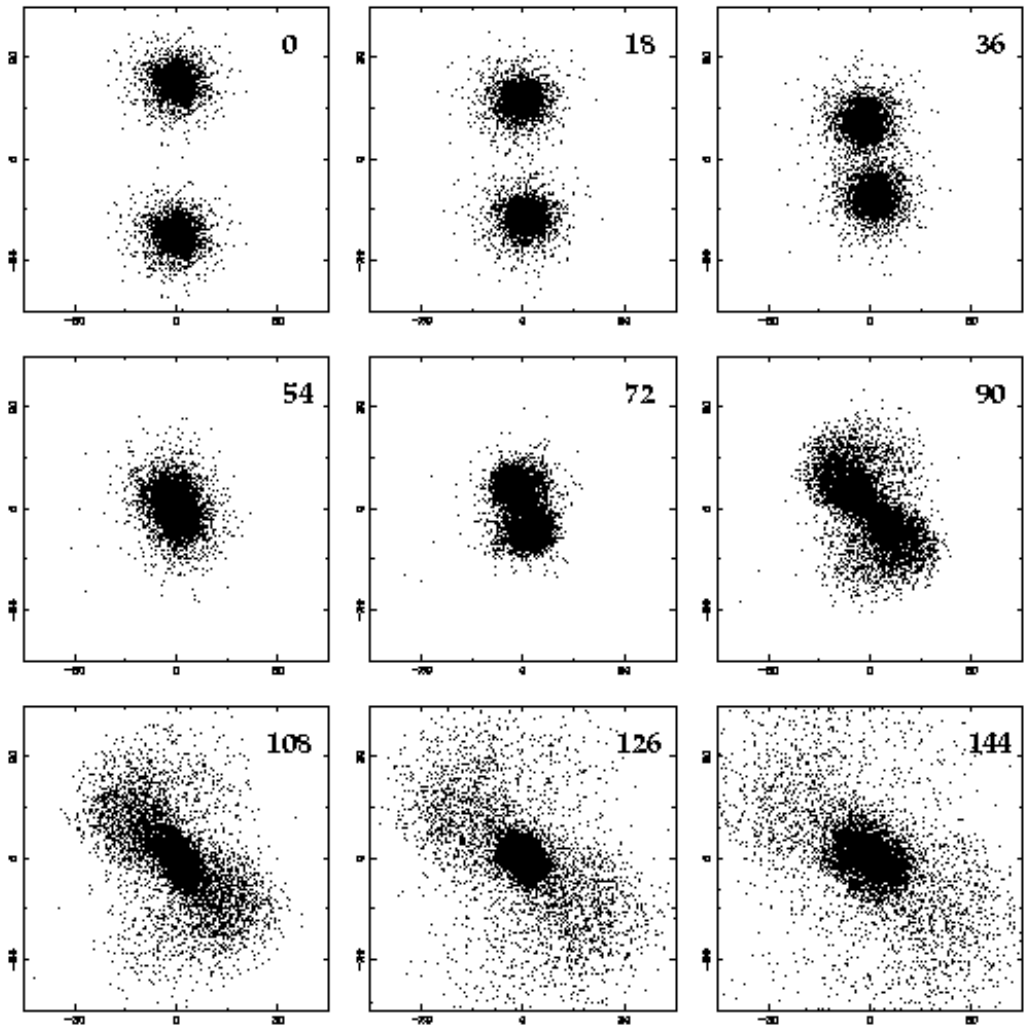}}
\caption{Evolution of systems in run $1o$. This is a collision between two equal mass galaxies on a parabolic orbit with a small impact parameter. Only luminous particles are shown. Numbers at the top of each frame show the time in computational units. The first encounter occurs around time 50.
 \label{fig:jh1_5evol}}

\end{center}
\end{figure*}

\begin{figure*}
\begin{center}
\leavevmode
\hbox{%
\epsfxsize=14cm
\epsfbox{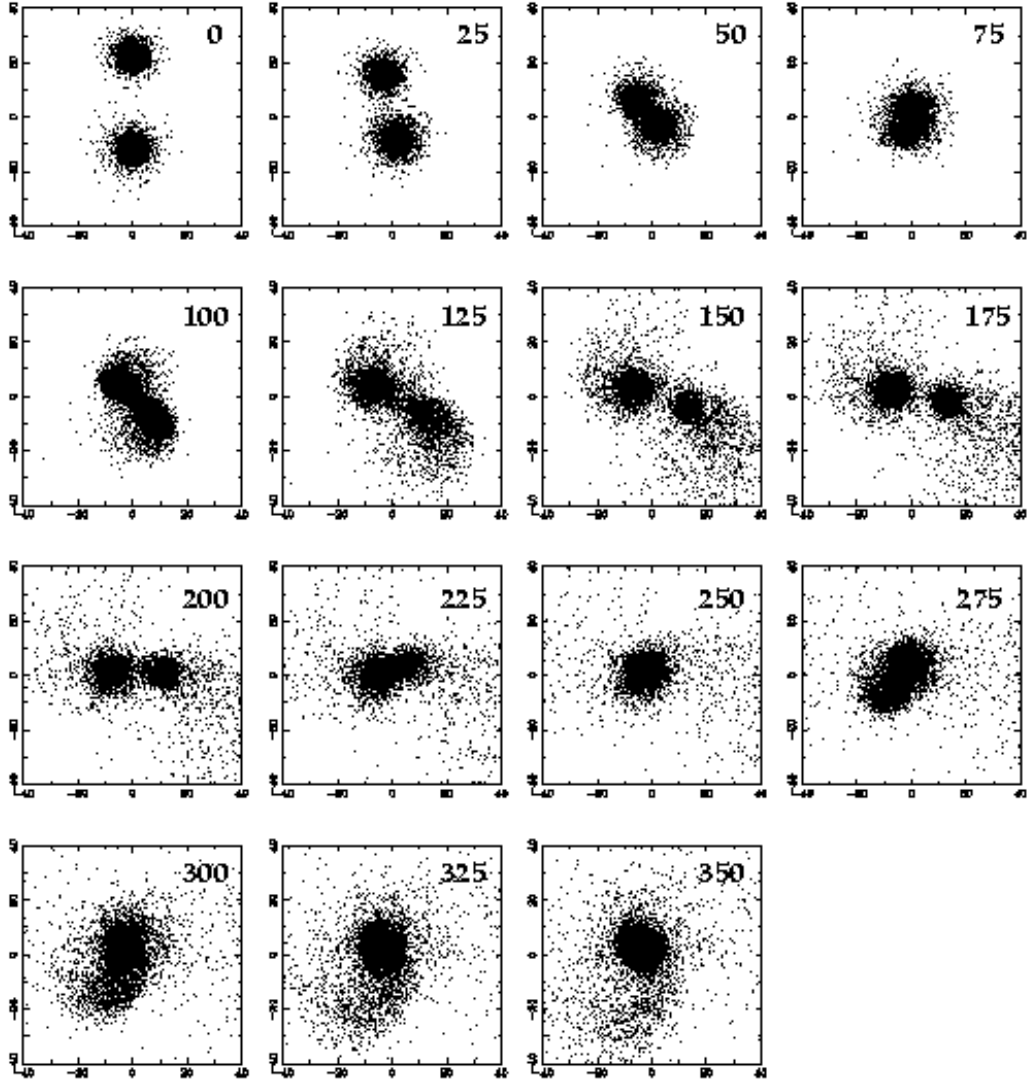}}
\caption{Evolution of systems in run $2g$. This is a collision between two galaxies with mass ratio 2:1 on a parabolic orbit with an impact parameter equal to the outer radius of the luminous part of the large galaxy. Only luminous particles are shown. Numbers at the top of each frame show the time in computational units. The first encounter is around time 60. Tidal tails or `plumes' form after the encounters, these consist mainly of material coming from the smallest system. \label{fig:jh2_10evol}}
\end{center}
\end{figure*}

\begin{figure*}
\begin{center}
\leavevmode
\hbox{%
\epsfxsize=14cm
\epsfbox{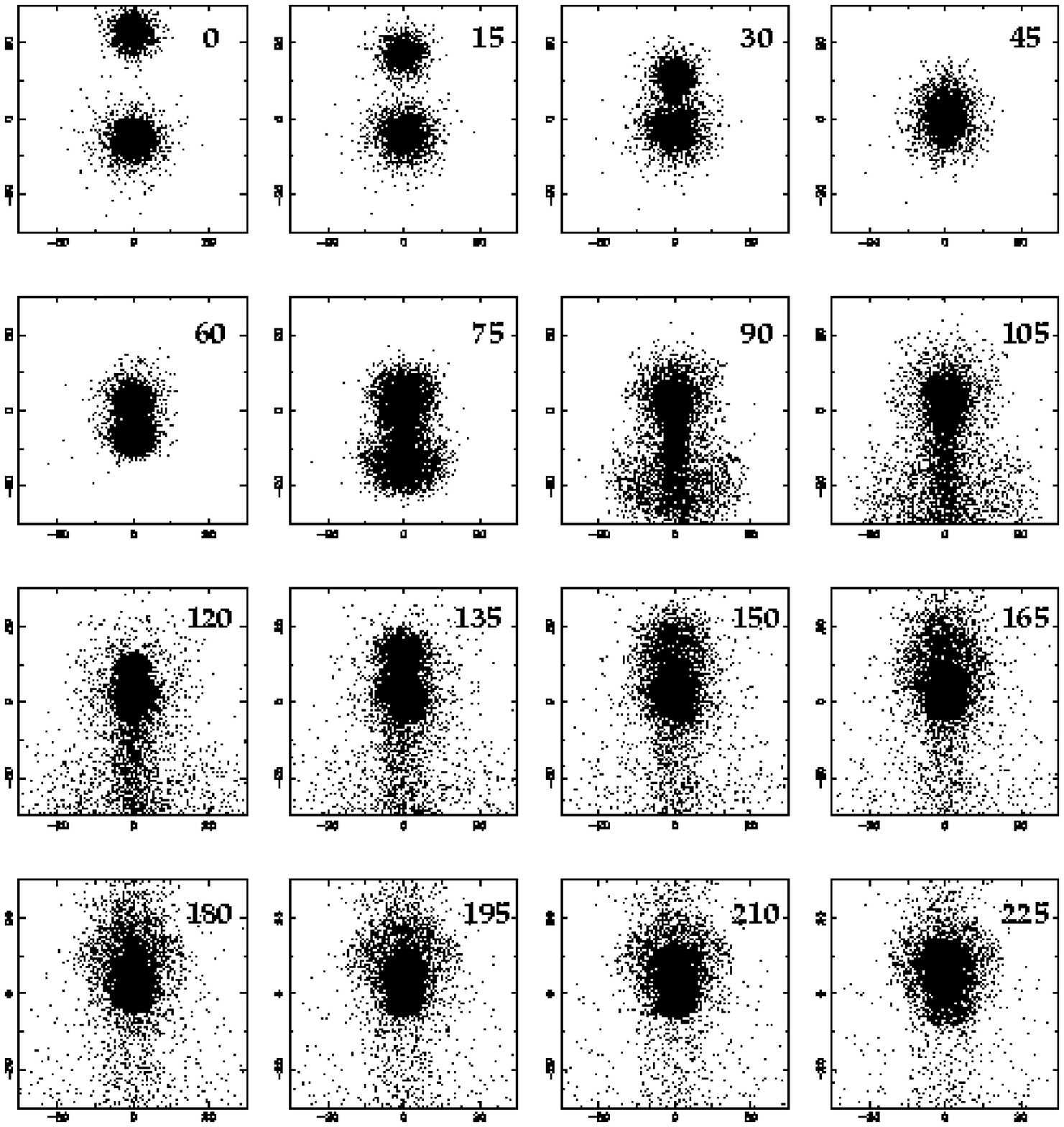}}
\caption{Evolution of systems in run $4h$. This is a head-on collision between two galaxies with mass ratio 4:1. Only luminous particles are shown. Numbers at the top of each frame show the time in computational units. The first encounter occurs around time 40. The particles from the smallest system are redistributed after several passages through the largest one and end up in prominent shells. \label{fig:jh4_0evol}}
\end{center}
\end{figure*}

In the present runs we have increased the total number of particles with respect to the simulations without halo by a factor of 5. The number of luminous particles, however, is kept low. We have proceeded in this manner due to the need to model live haloes with massive particles. If particles are too massive, particle-particle approaches are expected to heat the system if the difference in mass between particles is large. Models were let to evolve for at least $8$ to $10$ dynamical times of the merged system after merging to allow the system to relax (reach virialization). Conservation of energy is good in all the runs, better than $0.5$ per cent .

\section{Results}

The main characteristics of the merger remnants resulting from these encounters are summarized in table \ref{tab:halo}. A short description of the relevant features is given below.

\subsection{Phenomenology}

Our models develop a variety of characteristics during the collision stages. Some examples are shown in Figures \ref{fig:jh1_5evol}, \ref{fig:jh2_10evol} and \ref{fig:jh4_0evol}.

Figure \ref{fig:jh1_5evol} is an example of a collision with mass ratio 1:1.
The two galaxies are placed on a parabolic orbit with an impact parameter equal to half the outer radius of the luminous bulge. Around time 50 the systems meet for the fist time. After this first encounter both galaxies develop `plumes'. A merger follows when the two galaxies meet for the second time, in an almost rectilinear orbit. Thus the final system is prolate. In Figure \ref{fig:jh2_10evol} a pair with mass ratio 2:1 is shown. The impact parameter in this encounter is equal to the outer radius of the largest galaxy. The smaller galaxy (at top in the first frame of Figure  \ref{fig:jh2_10evol}) develops tails after the encounters around times 60 and 250. The final orbit, prior to the encounter is less rectilinear than in the previous case. The final system is triaxial.

Finally, Figure \ref{fig:jh4_0evol} shows a head-on collision for a run with mass ratio 4:1. The small galaxy (system at top in the first frame of Figure \ref{fig:jh4_0evol}) suffers a reorganization of its particles in a system of shells after several passages through the potential well of its massive neighbour. The shells are visible in the final frames; see also Figures~\ref{shelljh12} and~\ref{vrvsr}.

The general behaviour of these models is similar to that of the models without dark halo in Paper I. The main difference is in the sharpness of the shells that can be seen in models with a halo. 

\subsection{Morphology of the systems}

\begin{figure}
\begin{center}
\leavevmode
\epsfxsize=7cm
\epsfbox{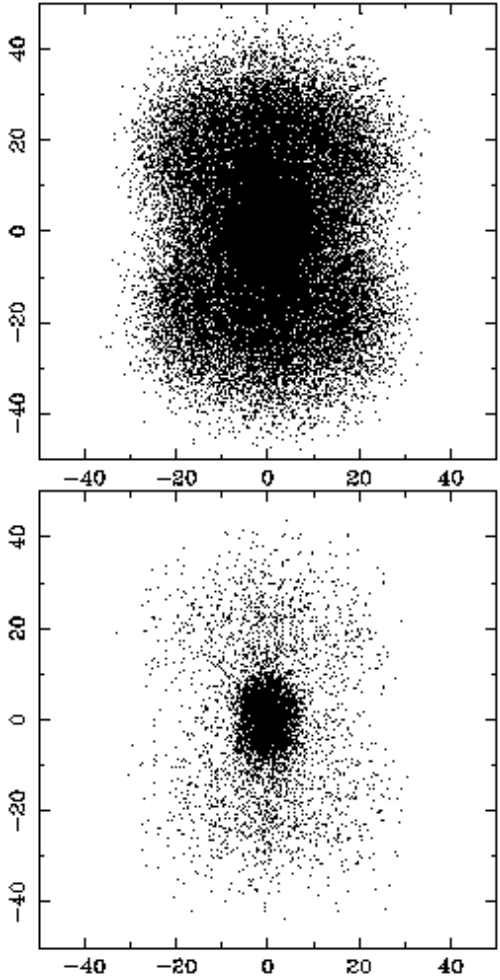}
\caption{Particle distributions for halo (top) and luminous (bottom) components for run $1h$ as seen along the z-axis. This is a head-on collision in a  parabolic orbit, which is reflected in the symmetry of the remnant. Some of the luminous particles in the plumes have enough energy to escape from the potential well, but most will fall back on the central system. \label{jh11}}
\end{center}
\end{figure}

\begin{figure}
\begin{center}
\leavevmode
\epsfxsize=7cm
\epsfbox{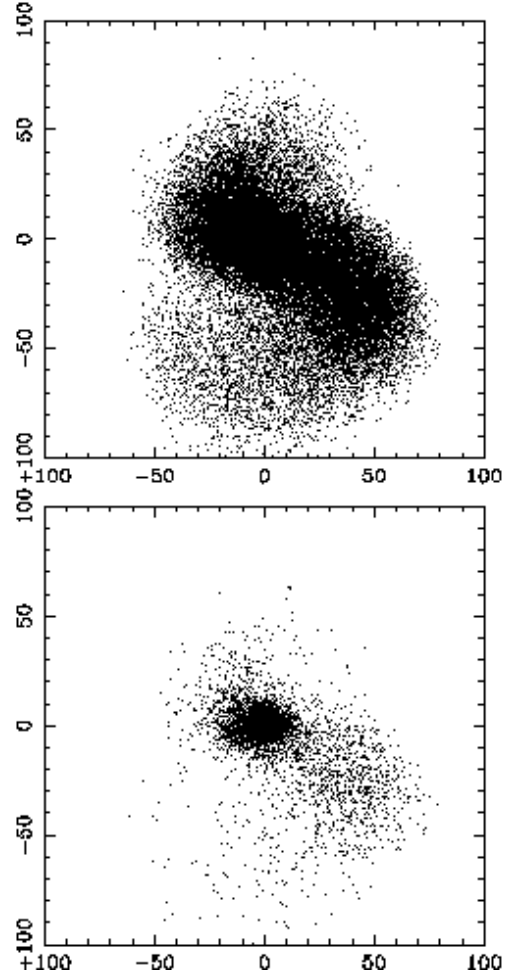}
\caption{Particle distributions for halo (top) and luminous (bottom) components for run $2o$ as seen along the z-axis. The initial configuration of this run has a parabolic orbit with a moderate impact parameter. This is reflected in the end product via the prominent tails in both the dark and luminous component. The large tail is coming from the progenitor of mass 1, i.e. the least massive galaxy.\label{jh12}}
\end{center}
\end{figure}

\begin{figure}
\begin{center}
\leavevmode
\epsfxsize=7cm
\epsfbox{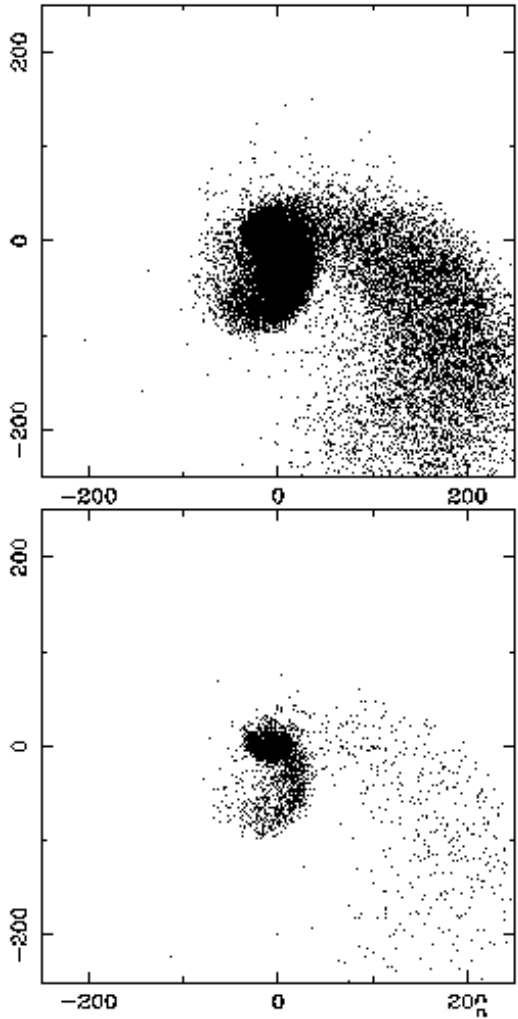}
\caption{Particle distributions for halo (top) and luminous (bottom) components for run $4g$ as seen along the z-axis. The initial configuration of this run has a parabolic orbit with a fairly large impact parameter. Several tails in both components can be seen. Tails are mainly coming from the system with smallest mass. In the inner parts prominent shells can be seen (not visible in this plot). All particles in the tails will eventually fall back to the inner regions since they do not have enough energy to escape.\label{jh14}}
\end{center}
\end{figure}

In Figures \ref{jh11}, \ref{jh12} and \ref{jh14} examples of the final states of different runs are shown. These illustrate the structural differences resulting from the variation in the orbital parameters. Each plot consists of two frames, the top one gives the distribution of halo particles, and the bottom one gives the distribution of luminous particles, both as seen along the z-axis, that is, the axis perpendicular to the orbital plane. 

In Figure \ref{jh11} the remnant of run $1h$ is shown.  This is a head-on parabolic collision, which is reflected in the symmetry of the remnant for both the dark and luminous component. Some of the luminous particles in the `plumes' have enough energy to escape from the potential well, but most will eventually fall back. Head-on collisions generally show a similar behaviour, with details depending on the masses of the progenitor systems. For non-equal-mass collisions, this symmetry seen in Figure \ref{jh11} is broken. The smaller system will be highly disrupted giving rise to a one-sided lobe in the halo.

Head-on collisions without dark matter show a similar behaviour. The end products for equal mass models also show plumes surrounding the main prolate core. The potential there is shallower however and a larger number of particles will eventually escape (see Paper I).

Figure \ref{jh12} shows the particle configuration for model $2o$. The initial (parabolic) orbit has a modest impact parameter. This is reflected in the end product via the prominent tails on both the dark (top panel) and the luminous (bottom) component. The larger one is coming from the progenitor of mass 1, which also presents small shells (see further below). Models with a moderate impact parameter usually merge after the second pass through the pericenter. Therefore, the merger remnant will have at least two or more tails, the precise number depending on the number of pericenter passages. 
The inner parts of the luminous component are nearly prolate. 

The end product of encounters of one-component systems with parabolic orbits (Paper I) is, in general, a triaxial spheroid. It also shows some evidence of tails developed during the interaction stages, and the number of tail is again connected to the number of pericenter passages. So, we find that the behaviour is similar.

Figure \ref{jh14} illustrates the final state of run $4g$. Here the parabolic orbit has a fairly large impact parameter. As a result, several tails in both components can be seen. Tails are mainly coming from the system with smallest mass at the beginning of the run. In the inner parts we could also see prominent shells like those already mentioned for models $2o$ and $4h$. As in those runs, these shells are formed from particles of the less massive system (see below). For simulations with a large impact parameter the tails are more prominent than for small $D$, because the galaxies merge only after several passes through the pericenter. Particles in the tails will eventually fall back to the inner regions because they do not have enough energy to escape. Note however that the simulations were stopped when the material in the tails was still falling back. The overall shape of the system will not be affected appreciably by the returning material since the mass involved in the tails is small. 
Since this material carries most of the angular momentum, it may change the kinematics of the inner parts.

Again, similar structures can be seen in simulations involving a fair amount of angular momentum in the sample without dark halo of Paper I. Several tails are formed as a consequence of the interactions and the exchange of orbital angular momentum. These tails or plumes carry away part of this angular momentum, that will eventually return to the main body when the particles in those tails fall back to the main body.

The dark haloes of the merger remnants have a variety of shapes depending on the initial conditions. For all our simulations, this shape is, grosso modo, the same as that of the luminous matter. Or, in other words, one can trace the shape of the dark matter using the luminous particles. 

Head-on equal mass merger collisions result in a prolate structure of the final luminous components. As expected, these are symmetric due to the initial symmetry introduced by the choice of orbital parameters. Equal mass mergers in general do show a high degree of symmetry both in the luminous and the dark matter. A similar symmetry results for head-on collisions with non-equal-mass mergers about the line connecting the initial systems. This would be reflected in a small isophote twisting for these systems depending on the point of view.

The asymmetries present in non-equal-mass off-axis mergers are more interesting. The smallest system is most affected by the encounter, forming prominent `plumes' or tidal tails while the large system remains almost undisturbed. These systems would show prominent isophote twisting as well.

These results are in good agreement with those in Paper I for similar initial mass ratios and orbital parameters.

\begin{figure}
\centering
\includegraphics[width=7cm]{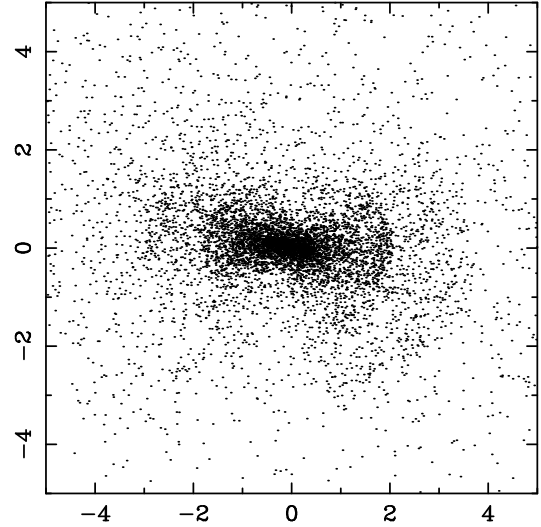}
\caption{Luminous particles from the small system in run $2o$. Many particles of this galaxy end up as shells in the merger remnant. \label{shelljh12}}
\end{figure}

\begin{figure}
\begin{center}
\leavevmode
\hbox{%
\epsfxsize=7cm
\epsfbox{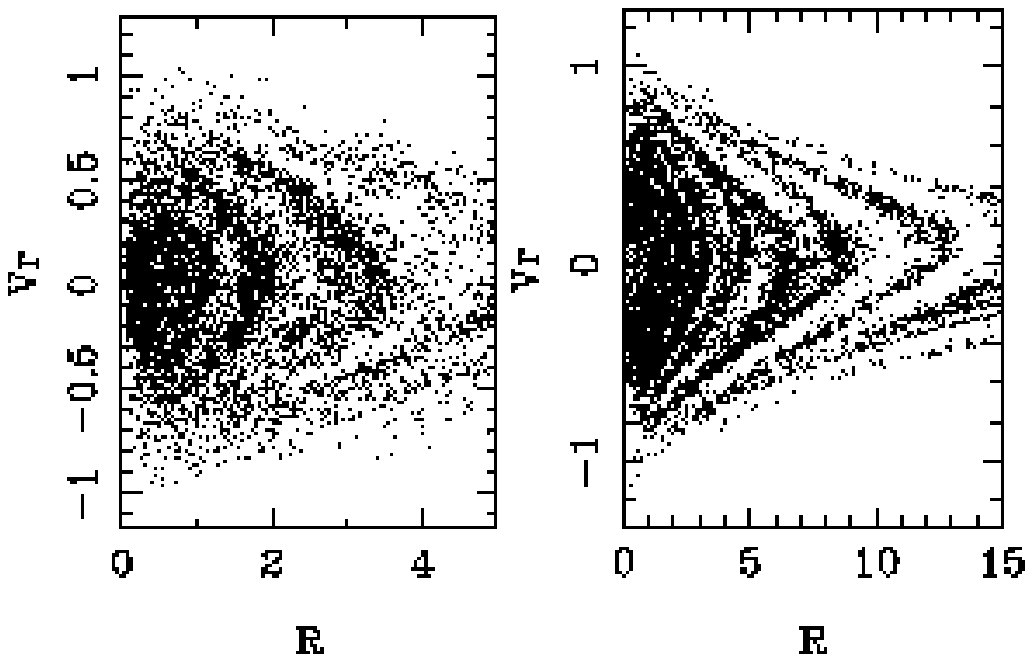}}
\caption {Radial velocity vs. radius diagrams ($V_{\rm r}$ vs. $R$) for the luminous particles of the smaller system for runs $2o$ (left panel) and $4h$ (right panel). The bands correspond to shells.\label{vrvsr}}
\end{center}
\end{figure}

Shells are among the prominent features we find in the non-equal-mass mergers in our sample. Figure \ref{shelljh12} shows the luminous particles of the small system in the final remnant of run $2o$. We show two examples of radial velocity versus spherical radius plots in Figure \ref{vrvsr} for runs $2o$ and $4h$. The first one is a parabolic encounter with a mild impact parameter. The shells look symmetric but are not very prominent. Run $4h$ is a head-on encounter where the particles of the small system end up in prominent shells.  

As stated in section 3.2 in Paper I, shells are also formed there. However, the shells found in models with a dark mater halo are more prominent than those found there. 

These features are summarized in table \ref{jhmorf} where a Hubble-type classification is given for the luminous part of the merger remnants. This classification is obtained after fitting the projected isodensity contours by ellipses for 100 random points of view. We have taken the mean projected ellipticity inside $R_{\rm e}$ (the radius enclosing half of the mass in projection). The classification given in table \ref{jhmorf} is the median of those ellipticities. (Note that in table \ref{tab:halo} we give the value $\epsilon$ of the ellipticity of the ellipse at $R_{\rm e}$ from the point perpendicual to the initial angula momentum vector). Also a description of the phenomenology in the halo and the luminous part of the remnant is given in table \ref{jhmorf}.

\subsection{Prolate and oblate systems}

\begin{figure}
\begin{center}
\leavevmode
\hbox{%
\epsfxsize=7cm
\epsfbox{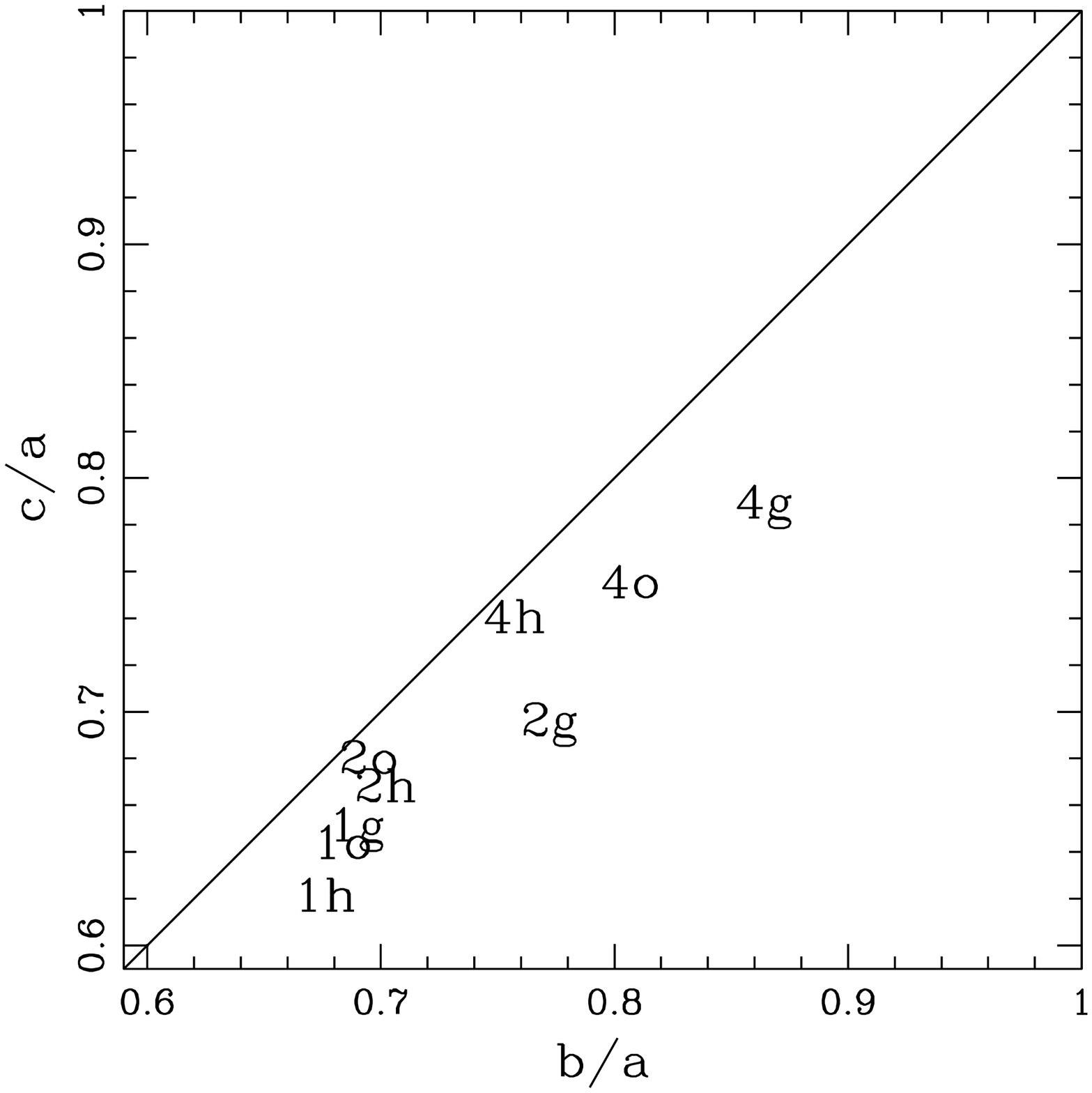}}
\caption{Following de Zeeuw \& Franx (1991), we have plotted the axis ratios of our merger remnants in $b/a$, $c/a$ space. Systems on the diagonal line are prolate, those on the line $b/a=1$ are oblate. \label{arathalo}}
\end{center}
\end{figure}

For each system we have measured the axial ratios $b/a$ and $c/a$, where $a$, $b$ and $c$ are the semi-axes of the luminous part, as calculated from the inertia tensor eigenvalues following the algorithm described in section 3.3 of Paper I. In these experiments with dark haloes the luminous remnants are mainly prolate or triaxial spheroids. The results are shown in Figure \ref{arathalo}. We fail to find oblate spheroids, but our sample is too limited to reach a firm conclusion regarding shapes.

When we compare with figure 10 in paper I, in which the non-halo models are plotted, we notice that in the simulations including haloes the final systems are more flattened than without halo. Values of $c/a$ as small as 0.6 are found, whereas without halo the most extreme flattening is 0.7. However there we did find oblate as well as triaxial and prolate spheroids. 

Because the choice for the ratio between dark and luminous mass has been rather arbitrary, one may speculate whether even more flattened and oblate systems could be found with the appropriate set of parameters.

\subsection{Rotation and flattening}

\begin{figure}
\begin{center}
\leavevmode
\hbox{%
\epsfxsize=8.5cm
\epsfbox{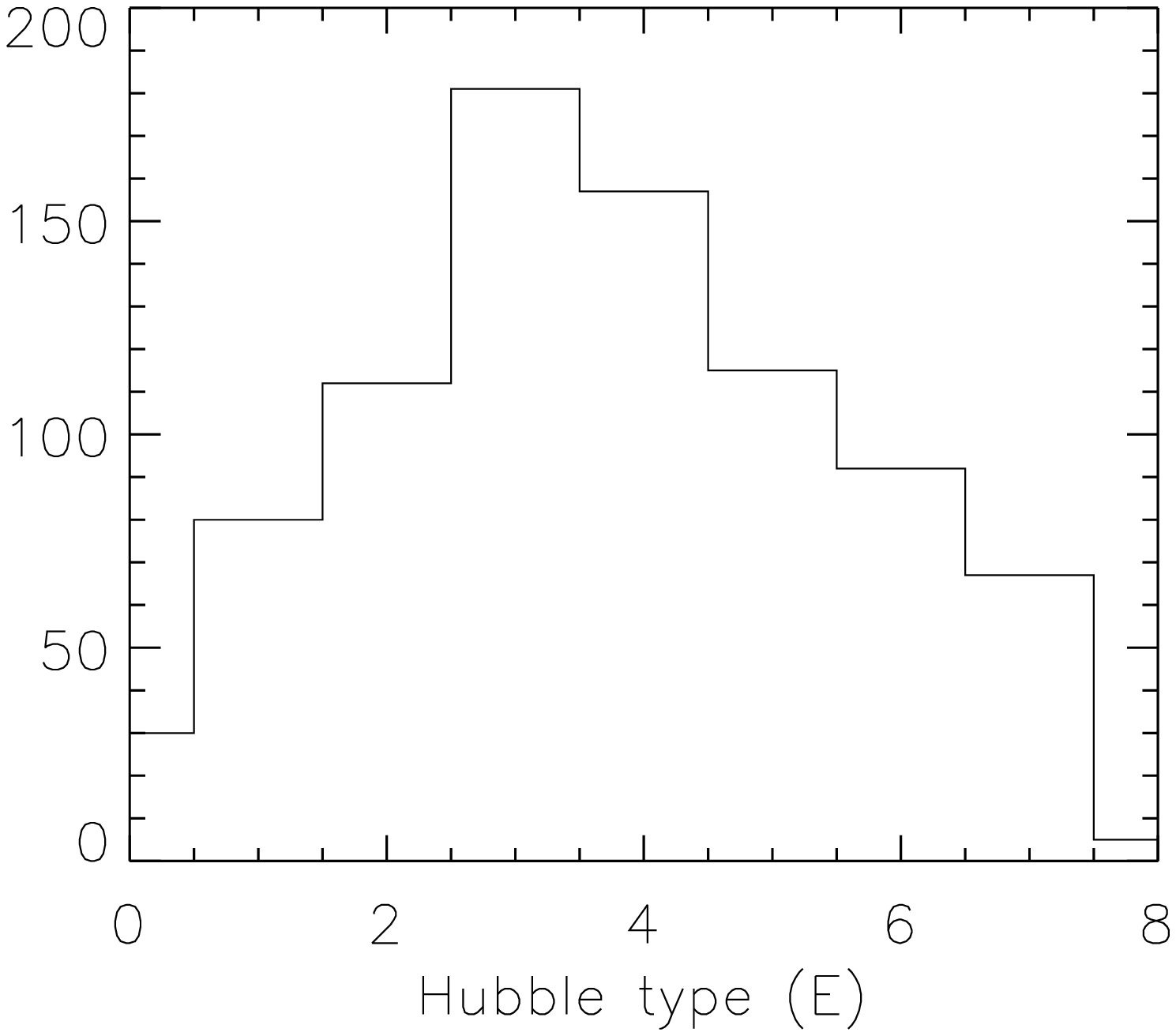}}
\caption{Histogram of Hubble types resulting from the merger simulations (all runs) by looking at the remnants from 100 random points of view. \label{typehalo}}
\end{center}
\end{figure}

\begin{figure}
\begin{center}
\leavevmode
\hbox{%
\epsfxsize=8cm
\epsfbox{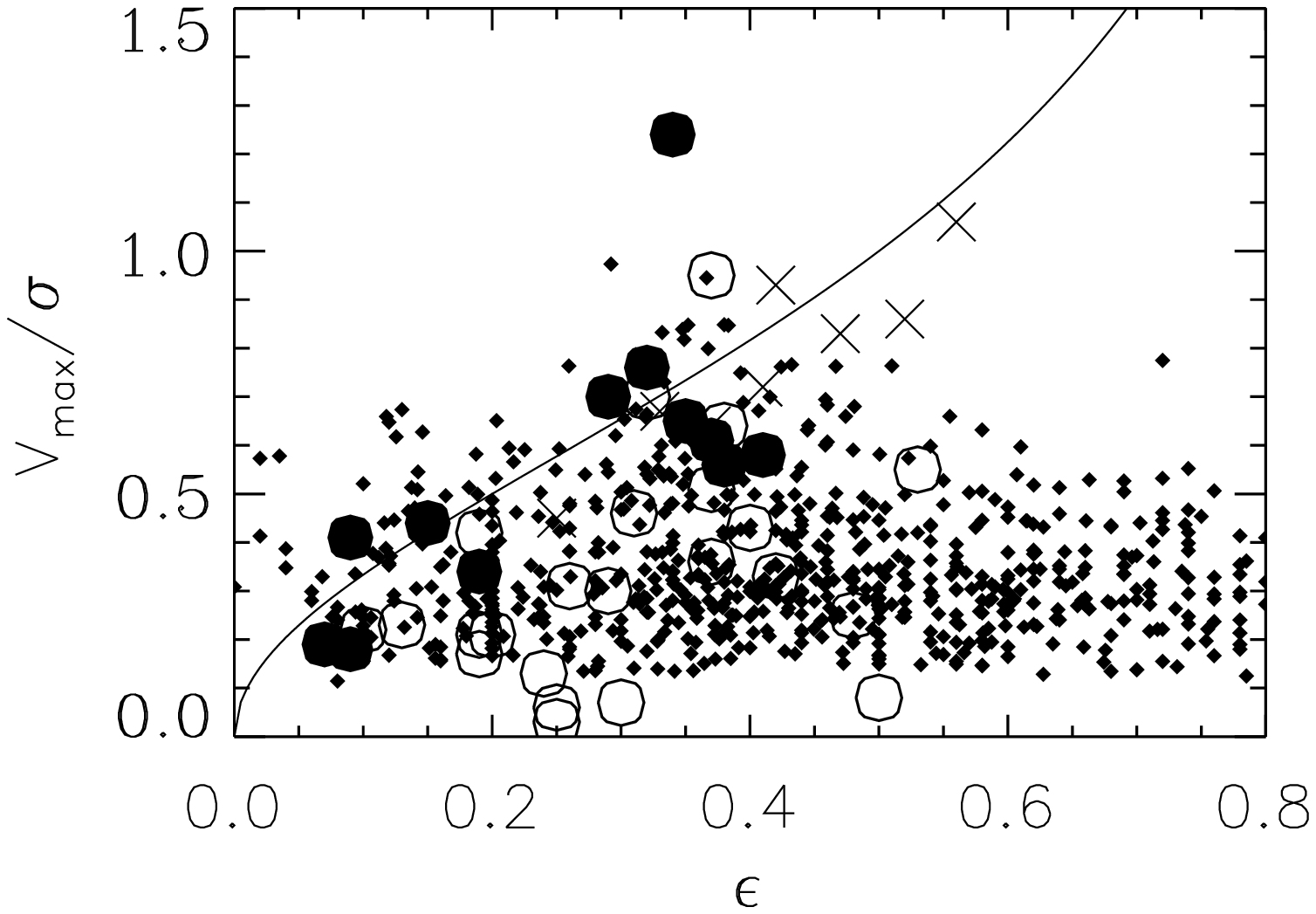}}
\hbox{%
\epsfxsize=8cm
\epsfbox{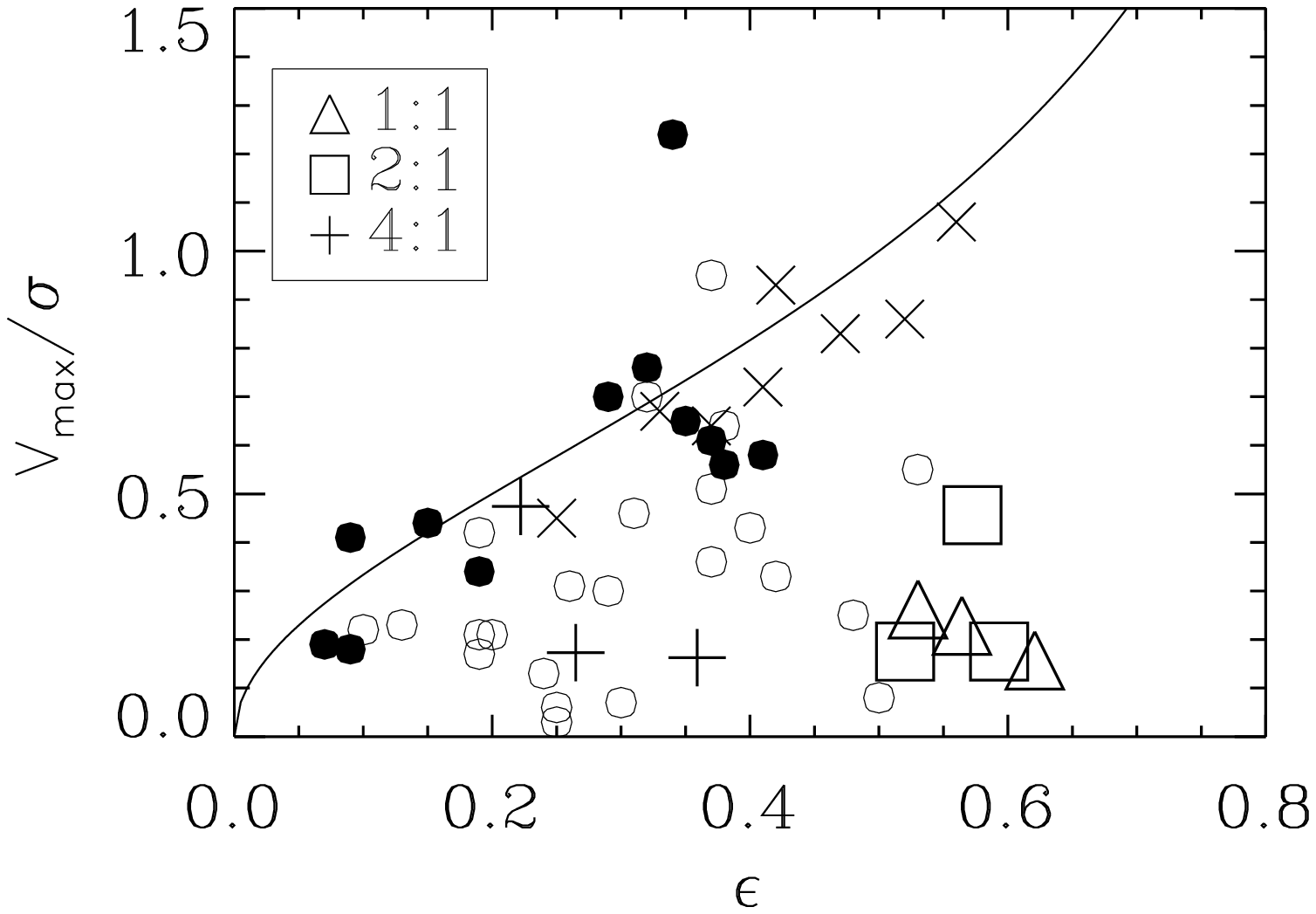}}
\caption{$V_{\rm max}/\sigma$ versus ellipticity diagram.  In both panels observational data from Davies et al. (1983) are plotted: open circles are high-luminosity ellipticals, filled circles are low-luminosity ellipticals and crosses are bulges. The upper panel gives the cloud of points obtained when looking at each model from one hundred points of view (small dots). The lower panel represents the models when seen from a point of view along the y-axis. Different symbols refer to the various mass-ratio's used; see inset. Compare with figure 11 in Paper I.  \label{evsehalo}}
\end{center}
\end{figure}

For each of our merger remnants we have fitted the projected isodensity contours by ellipses for 100 random points of view. Then we have measured the ellipticity for the luminous part inside the half luminous-mass radius. 
The mean of these results are shown in Figure \ref{typehalo}.
The full range of morphological types, from E0 to E7 is covered. It may come as a surprise that E6/E7 systems are produced in collisionless N-body simulations. We have therefore checked this result by visual inspection of the isophote shapes in a number of cases. Although the isophotes are sometimes rather irregular, our visual inspection confirms the E6/E7 shape inside about one effective radius. Farther out the isophotes often have a peanut shape. It appears that in these cases one is looking at the end result of a head-on collision from a point of view perpendicular to the relative orbit. 

We find a peak at E3, as is observed in real-life ellipticals, although this is most probably accidental, given  the restricted set of parameters used. As found in the simulations without dark halo, the runs in which the progenitors are of comparable mass result in higher ellipticities.

In Figure \ref{evsehalo}, $V_{\rm max}/\sigma_{\rm o}$ is plotted versus $\epsilon$. Over-plotted are observational data from Davies et al.  (1983). We find that the models can cover most of the observed range. In general we see that the maximum rotation is not high. For a model with a large mass ratio (4:1) the merger process is still effective in forming a rotating system.
When comparing this with figure 11 in Paper I we see that both samples are able to reproduce mildly rotating ellipticals. We find a difference, though, for the fast rotators. For non-halo models we find that models with a large amount of angular momentum lie close to the locus of the oblate rotators. For models with halo we only find this for models with a large mass ratio and large angular momentum. However we do not have flattened rotating oblate spheroids in the sample with dark halo.

These results indicate that the amount of dark matter (together with the orbital angular momentum and mass ratios, see Paper I) determines the amount of rotation, i.e. angular momentum, that is transfered to the luminous parts.

\subsection{Boxiness-disciness}

We have calculated the deviation from pure ellipses for our merger simulations.

When a deviation at a given isophotal level is discy the $a_4$ parameter is positive while for boxy deviations this parameter is negative (see Binney \& Merrifield 1998, and section 3.5 in Paper I).

\begin{figure}
\begin{center}
\leavevmode
\hbox{%
\epsfxsize=7cm
\epsfbox{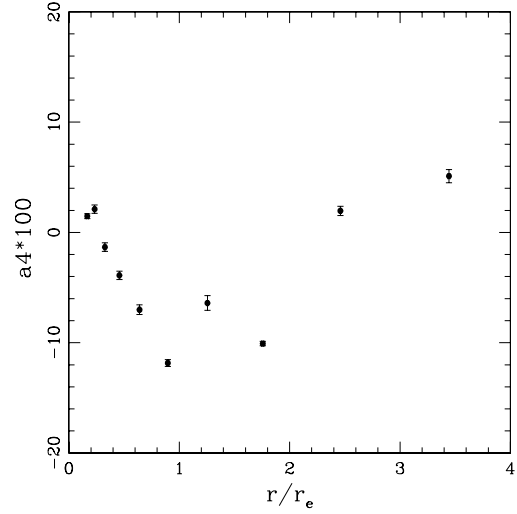}}
\caption{Mean radial variation of $a_4$ after equilibrium has been reached for model $1g$. Negative $a_4$ indicates boxy isophotes. \label{boxihalo}}
\end{center}
\end{figure}

In Figure \ref{boxihalo} we present the deviations for model $1g$.
To increase the signal to noise ratio an average over 60 snapshots was made, always calculating the isophotes for a projection parallel to the intermediate-axis. These 60 snapshots are obtained by evolving the end products a bit further in time.

We find that a highly boxy system is formed. The remnant is boxy out to two effective radii.
The large boxiness found here is in contrast with the modest boxiness found in the models without halo. Here we find that all systems are boxy while some of the systems in Paper I, using similar orbital parametrers, are discy.

\subsection{Kinematics}

For the models with haloes we do not find such a clear rotation as found for non-halo mergers (see section 3.6 in Paper I). 
Only weak rotation is found in the cases with the highest impact parameter (larger angular momentum).

\begin{figure}
\begin{center}
\leavevmode
\hbox{%
\epsfxsize=9cm
\epsfbox{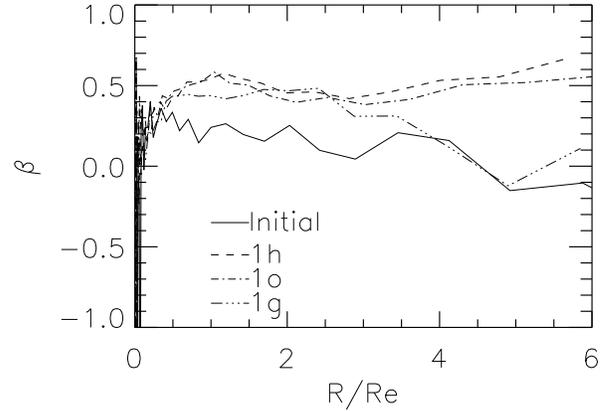}}
\caption{Anisotropy parameter versus radius. Positive values indicate radial anisotropy, negative ones indicate tangential anisotropy.\label{anisohalo}}
\end{center}
\end{figure}

Elliptical galaxies are mainly supported by random motions. To measure the velocity anisotropy we used the $\beta$ parameter defined as:

\begin{equation}
\beta= 1 - \frac{\sigma^2_{\rm t}}{2 \sigma^2_{\rm r}},
\end{equation}
with $\sigma_{\rm t}$ the tangential velocity dispersion ($\sigma_{\rm t}^2=\sigma_{\rm \theta}^2 + \sigma_{\rm \phi}^2$) and $\sigma_{\rm r}$ the radial velocity dispersion; see section 3.6 in Paper I.

Figure \ref{anisohalo} shows how this parameter changes with radius, for the luminous matter only. Four different models are plotted. The initial relaxed system is plotted as a solid line. As mentioned above, although initially isotropic, after relaxation the inner parts of the system develop some radial anisotropy.
Dotted-dashed lines show different end results of our simulations. All refer to models with the same mass ratio but different orbital parameters. Up to a radius close to $R=4R_{\rm e}$ all systems are radially anisotropic (where $R_{\rm e}$ is the radius that includes half of the luminous mass in projection). For model $1h$, with the largest impact parameter, the outer parts are close to isotropic. All merger remnants are more radially anisotropic than our initial relaxed system.

Merger remnants without halo resulting from head-on collisions also show radial anisotropy, while models with $D \neq 0$ show a trend that the tangential anisotropy of the remnants increases with increasing impact parameter. The fact that this trend is not clear in the present sample is probably related to the radial anisotropy that the initial system develops and to the effect of the halo in absorbing a significant fraction of the angular momentum.

\section{Discussion}

We have carried out simulations of collisions between two-component realizations of elliptical galaxies consisting of a luminous and a dark halo component. The initial parameters are chosen such that all simulations end up in mergers.

The luminous parts of the merger remnants show a tendency to be prolate or triaxial, regardless of the initial angular momentum content of the orbit. This appears to be the case, because at the final stages prior to merger the systems encounter each other in nearly rectilinear orbits. Apparently most of the orbital angular momentum is absorbed by the halo particles. Most of the luminous particles have radial orbits, giving the final cigar-shape figure. Since we start with spherical systems it comes as no surprise that the remnants are closest to a sphere when the masses of the initial systems are dissimilar, as is also the case for models without halo.

This prolate to triaxial general shape is in agreement with claims by Ryden (1996), although Lambas, Maddox \& Loveday  (1992) argue that a triaxial shape is more consistent with the observed ellipticities in elliptical galaxies. Alam \& Ryden (2002) find from the SLOAN digital sky survey that they can rule out with $99$ per cent  confidence that elliptical galaxies are oblate spheroids. Vincent \& Ryden (2005) exclude with a 99 \% confidence level that de Vaucouleurs galaxies from SLOAN Data Release 3 are oblate systems with equal tri-axiality parameter for all isophotes. Bak \& Slater (2000) on the contrary find a bimodal distribution with oblate and prolate spheroids.

Our systems show a peak in their Hubble type distribution near E3. The sample is by no means complete. However, all remnants show the tendency to be prolate or triaxial and this result seems robust. Lambas et al. (1992) show that an E3 Hubble-type is favored. Similarly, Franx, Illingworth \& de Zeeuw (1991) find that there is a peak around E3 in real elliptical galaxies. The situation is less clear in models without halo. In Paper I we stressed that our results with regard to the final morphology must be put in the frame work of the likelihood of having the specific initial conditions that we used. 

Our remnants from systems with halo show low rotation. In general our merger remnants are non-rotating. This is also found in Paper I for head-on collisions. But here it is true for head-on as well as non-head-on collisions. For the encounters involving a non-zero angular momentum in models without halo (Paper I) we found that the orbital angular momentum is transfered to spin angular momentum. For the present two-component models this is also true. However, now it is the halo which retains the spin angular momentum. As a result this leaves the particles in the luminous bulges in radial orbits right before the final encounter leading to the merger. This final encounter is a nearly head-on collision, no transfer of angular momentum to the bulges takes place, and the final system takes a nearly prolate shape. 

This final encounter, almost head-on, is also responsible for the radial anisotropy found in the inner parts of the merger remnants in most of our simulations. The outer parts of the remnants from the simulations with large orbital angular momentum gain a small amount of rotation and are close to isotropic.
In general this would be in good agreement with data for high-luminosity elliptical galaxies. Therefore we might propose mergers between spheroids with a halo as a possible origin for those systems.

Although the amount of dark matter in elliptical galaxies is still uncertain (Baes \& Dejonghe 2001, Romanowsky et al. 2003), it is of interest to look at the effect of dark matter on the properties of the merger remnants and the merging process in general.
We can therefore compare the two samples, the one presented here and the one described in Paper I in a more general sense.

Head-on collisions for equal mass mergers show a very similar behaviour in both samples. The final systems are prolate, non-rotating radially anisotropic spheroids.
Equal mass mergers with $D \neq 0$ show some differences. Non-halo models result in rotating spheroids supported by tangential anisotropy and may develop a bar or discy isophotes depending on the mass ratio of the progenitors. Halo models produce nearly prolate, radially anisotropic spheroids, mainly with boxy isophotes.

In this regard, we could trace the differences in shape and anisotropy observed in real elliptical galaxies to different formation mechanisms and contents of dark matter.

The results from Saglia et al. (1993) and Capaccioli et al. (2002) show that there might be two different populations of elliptical galaxies with regard to their dark matter halo. While elliptical galaxies with a dark halo show boxy deviations in their isophotes, galaxies that show no evidence for dark matter tend to be more discy. This difference may tell us something about different formation mechanisms. 

The simulations in Paper I refer to mergers of spheroids without a dark matter halo. This situation may apply to spherical systems surrounded by the global potential well of a cluster. Two cases may apply here. Elliptical galaxies in the central region of the cluster may have lost their halo, which is now part of the cluster halo. Another possibility is that the global potential well of the cluster dominates.
The case {\it with} dark matter haloes would seem to be more relevant for elliptical galaxies in the field.

High-resolution cosmological N-body simulations produce cluster halos containing many `subhaloes' which are plausibly identified as the dark haloes of individual galaxies (Moore et al.~1999, Klyping et al.~1999). These haloes, although truncated, may still be massive and extended enough to play a dynamical role in E+E mergers, blurring somehow the above scheme.

Such a scheme could in principle be easily tested with observations given the differences in morphology and kinematics observed in our two samples. But the possibility that elliptical galaxies can be built out of mergers of discs introduces a complication. 

Based on the simulations just described, and according to the scheme presented above, we can propose a formation mechanism for these two different populations of elliptical galaxies. Those with signatures of a dark matter halo might be formed by a merger of two spheroidal systems surrounded by a halo, probably in the field or in compact groups (Kelm \& Focardi, 2004). The remnant would show preferentially boxy deviations. Those without a clear signature of a halo might be formed inside a cluster and the formation mechanism followed would be that described by the models in Paper I, with a broad range of different characteristics and where some of them have discy deviations in their isophotes. In other words, we could expect some habitat segregation but opposite to the one reported by Shioya \& Taniguchi (1993). This calls for further observational work.

Given the large range in properties of the merger remnants in our simulations we may conclude that the formation of elliptical galaxies (at least for the high-luminosity ones) can in principle be explained by mergers of systems dominated by spheroidal components.

\section{acknowledgments}
We thank the referee for his/her constructive comments.

\end{document}